\documentclass[sigconf]{acmart}
\usepackage{comment}
\usepackage{multirow}
\usepackage{booktabs}
\AtBeginDocument{%
  \providecommand\BibTeX{{%
    \normalfont B\kern-0.5em{\scshape i\kern-0.25em b}\kern-0.8em\TeX}}}

\copyrightyear{2020}
\acmYear{2020}
\acmConference[WWW '20]{Proceedings of The Web Conference 2020}{April 20--24, 2020}{Taipei, Taiwan}
\acmBooktitle{Proceedings of The Web Conference 2020 (WWW '20), April 20--24, 2020, Taipei, Taiwan}
\acmPrice{}
\acmDOI{10.1145/3366423.3380266}
\acmISBN{978-1-4503-7023-3/20/04}

\begin{document}

\title{Learning to Hash with Graph Neural Networks \\ for Recommender Systems}

\author{Qiaoyu Tan, Ninghao Liu, Xing Zhao}
\authornotemark[1]
\affiliation{%
  \institution{Texas A\&M University}
}
\email{{qytan,nhliu43,xingzhao}@tamu.edu}

\author{Hongxia Yang, Jingren Zhou}
\authornotemark[2]
\affiliation{%
  \institution{Alibaba Group}
}
\email{{yang.yhx, jingren.zhou}@alibaba-inc.com}

\author{Xia Hu}
\authornotemark[1]
\affiliation{%
  \institution{Texas A\&M University}
}
\email{xiahu@tamu.edu}
\renewcommand{\shortauthors}{Qiaoyu Tan, et al.}

\begin{abstract}
Recommender systems in industry generally include two stages: recall and ranking. Recall refers to efficiently identify hundreds of candidate items that user may interest in from a large volume of item corpus, while the latter aims to output a precise ranking list using complex ranking models. Recently,
graph representation learning has attracted much attention in supporting high quality candidate search at scale. Despite its effectiveness in learning embedding vectors for objects in the user-item interaction network, the computational costs to infer users' preferences in continuous embedding space are tremendous. 
In this work, we investigate the problem of hashing with graph neural networks (GNNs) for high quality retrieval, and propose a simple yet effective discrete representation learning framework to jointly learn continuous and discrete codes. 
Specifically, a deep hashing with GNNs (HashGNN) is presented, which consists of two components, a GNN encoder for learning node representations, and a hash layer for encoding representations to hash codes. The whole architecture is trained end-to-end by jointly optimizing two losses, i.e., reconstruction loss from reconstructing observed links, and ranking loss from preserving the relative ordering of hash codes. A novel discrete optimization strategy based on straight through estimator (STE) with guidance is proposed. The principal idea is to avoid gradient magnification in back-propagation of STE with continuous embedding guidance, in which we begin from learning an easier network that mimic the continuous embedding and let it evolve during the training until it finally goes back to STE. Comprehensive experiments over three publicly available and one real-world Alibaba company datasets demonstrate that our model not only can achieve comparable performance compared with its continuous counterpart but also runs multiple times faster during inference.    

\end{abstract}



\maketitle

\section{Introduction}
Recommender system has become the fundamental tool in our daily life to support various services online, i.e., web search and E-commerce platforms. Given a query, the recommendation engine is expected to recommend a small set of items that users prefer from the database. In practice, this process contains two important stages: Recall and Ranking. Recall process aims to efficiently retrieval hundreds of candidate items from the source corpus, e.g., million items, while ranking refers to generate a accurate ranking list using predictive ranking models. 
Currently, network embedding approach has been extensively studied in recommendation scenarios to improve the recall quality at scale. These methods aim to represent each object in the user-item interaction network with a low-dimensional continuous embedding vector $\textbf{z}\in \mathbb{R}^d$, expecting that similar users or items have similar embeddings. Among them, graph neural networks (GNNs)~\cite{kipf2016semi,hamilton2017inductive}, as a special instantiation of neural networks for structured data, have achieved state-of-the-art performance in information retrieval~\cite{fan2019graph}. Despite their retrieval quality, the computational costs to filter such a number of candidates in continuous embedding space are expensive, due to the inference inefficiency with $O(Nd)$ computational complexity for linear search in the worse case, where $N$ is the total number of objects in corpus. Therefore, it's a promising direction to improve the efficiency of recall stage in real-world recommendation.

\begin{figure}[h]
  \centering
  \includegraphics[width=8.5cm,height=6.5cm]{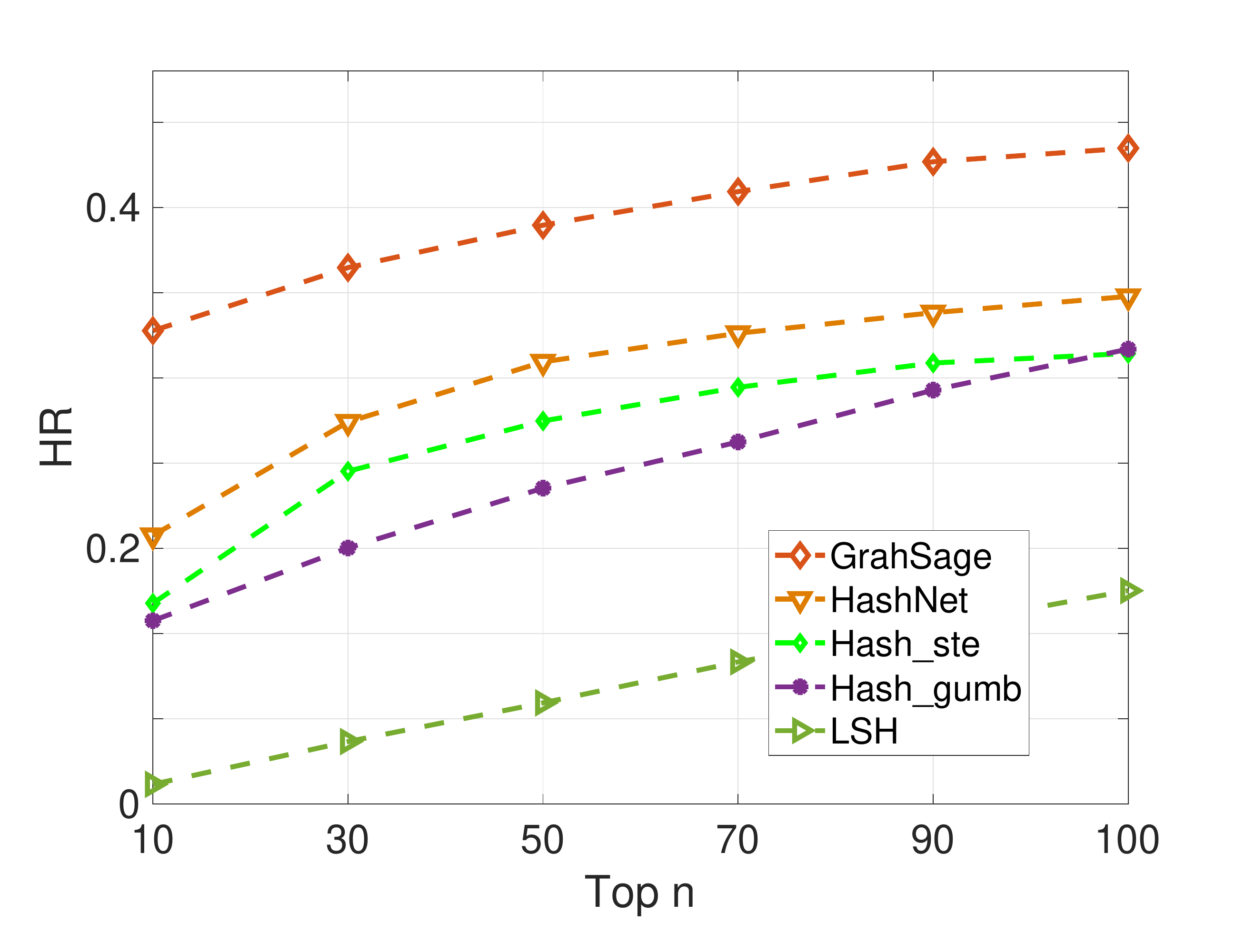}
\caption{Comparisons between GraphSage~\cite{hamilton2017inductive} and state-of-the-art hashing methods on Gowalla. We can observe that GraphSage performs significantly better than others on recommendation scenarios. It indicates the gap between continuous embedding and corresponding binary hash code. }
  \label{figure1}
\end{figure}

To overcome the computational barriers, hashing~\cite{wang2015learning} has attracted increasing attention due to its great efficiency in retrieving from large data. The basic idea of hashing is to construct a mapping functions to index each object into a compact binary code, minimizing the Hamming distances for similar objects and maximizing on dissimilar ones. 
Recently, hand-craft feature based models \cite{gionis1999similarity,gong2012iterative,liu2012supervised} and deep models~\cite{lai2015simultaneous,erin2015deep,zhu2016deep,liu2016deep} have been proposed, where the formers seek
hashing function on hand-crafted features and separate the encoding of feature representations from their quantization to hashing codes, resulting in sub-optimal solutions. The latters jointly learn feature representations and hashing projections. While encouraging performances are reported in these methods, the key disadvantage of them is that they need to first learn continuous representations which are then binarized into hashing codes in a separated post-step with sign thresholding. Therefore, with continuous relaxation, these methods essentially solve an optimization problem that deviates from the hashing objectives, and may fail to generate compact binary codes. 
Besides, for those deep hashing methods that are mainly tailored to generate high quality binary codes, the representative ability of associated continuous embeddings may be poor as shown in Figure~\ref{figure1}. We can observe that the retrieval accuracy of state-of-the-art hashing methods are inferior to the corresponding continuous embeddings in recommendation scenarios. As a result, it impedes the implementation of finer-level ranking in recommender systems. 

To bridge the gap, we propose an end-to-end discrete representation learning framework that can optimize over both continuous and binary representations, which embraces two challenges. 
First, to accurately generate binary hash codes $\mathbf{h}$ from associated continuous embeddings $\mathbf{z}$, we need to adopt the $sign$ function $\mathbf{h}=sign(\mathbf{z})$ as activation on the top of hashing layer. However, the gradient of $sign$ function is zero for all nonzero values in $\mathbf{z}$, which makes standard back-propagation infeasible, resulting in $\mathbf{h}$ and $\mathbf{z}$ not compatible. Second, unlike classification problem, in recommendation scenarios we care more relative ranking ordering. Therefore, training losses that merely focus on reconstruction probability of observed link~\cite{hamilton2017inductive} or semantic label information~\cite{kipf2016semi} will fail to preserve the ranking structure in hashing codes, making hamming space retrieval ineffective. Optimizing deep hashing networks with sign activation that could output high quality binary hashing codes as well as satisfactory continuous embeddings remains an open problem for learning to hash.

To address the challenges above, we propose a novel end-to-end learning framework for hashing graph data, named HashGNN. The developed model is generally applicable to arbitrary GNNs. The learned hash codes could improve the efficiency of item retrieval, while the associated continuous representations could improve the capacity of ranking model in modern recommender system. The whole architecture is optimized towards a task-specific loss function that not only preserves the topology of input graph but also maintains its ranking structure in the hamming space. To enable truly end-to-end training, we adapt a novel discrete optimization strategy based on straight through estimator (STE)~\cite{bengio2013estimating} with guidance. We first argue that STE may incur noisy gradients in back-propagation by magnifying the gradient of corresponding continuous embedding based optimization. Then we introduce a guidance approach to eliminate the gradient noise of STE, and turn it into a different problem that is easier to optimize. The contribution of this paper is summarized as follows:
\begin{itemize}
\item We tackle the problem of unsupervised hashing with GNNs trained in a end-to-end manner, in which both high quality hashing codes and continuous embeddings are returned. To the best of our knowledge, this paper represents the first effort towards this target in recommendation.

\item A simple discrete optimization strategy is adapted to optimize the parameters, which facilitates both efficient and effective retrieval in downstream tasks. 

\item Extensive experiments on four real-world datasets of different volumes demonstrate the advantages of our proposal over several state-of-the-art hashing techniques.

\end{itemize}

\section{Related Work}
We briefly group the related works into two categories: graph representation learning, learning to hash. The former draws upon research in encoding objects in graph into continuous representations, while the latter investigates encoding objects into binary hash codes.

\textbf{Graph representation learning.} This line of work has been studied and analyzed widely in social network analysis and recommendation scenarios~\cite{hamilton2017representation,zhang2019deep,tan2019deep}. Representatives including models based on matrix factorization~\cite{tang2015line,qiu2018network,liu2019single}, random walks~\cite{perozzi2014deepwalk,dong2017metapath2vec,grover2016node2vec} or deep learning~\cite{cao2016deep, wang2016structural,velivckovic2017graph}. Among them, graph neural networks, as a special instantiation of convolutional neural networks for structured data, has received a lot of attention for its great power in generating embedding~\cite{kipf2016semi,ying2018hierarchical} as well as scalability for large-scale graphs~\cite{ying2018graph}. 
Although the aforementioned methods have been proved to be very effective in generating embeddings, they are sufferred from inference inefficient due to the high computational cost of similarity calculation between continuous embeddings.

\textbf{Learning to hash.} Researches in this direction has proceeded along two dimensions: unsupervised hashing and supervised hashing~\cite{wang2017survey}. Unsupervised hashing methods~\cite{gong2012iterative,liu2011hashing,jegou2010product,salakhutdinov2007learning} aim to learn hash functions that encode objects to binary codes by training from unlabeled data, while supervised hash methods~\cite{kulis2009learning,liu2012supervised,norouzi2011minimal,shen2015supervised} target to explore supervised signals to help in generating more discriminative hash codes. Thanks to the development of deep learning, researchers also resort to blend the power of deep learning for hash coding~\cite{zhu2016deep,xia2014supervised,lai2015simultaneous,lin2015deep,shen2015supervised,erin2015deep} recently. Although encouraging performance are reported in the these deep hash methods, they all need to first learn continuous deep representations, and then generate the exactly binary hash codes by binarizing the learned representations with a separated post-step, which is actually deviates significantly from the original hash objects, thus may result in substantial loss of retrieval quality~\cite{gong2012iterative}. A recent work named HashNet~\cite{cao2017hashnet} proposes to lower the quantization error of binarization by continuation technique. It approximates the $sign$ function with scaled tanh function $tanh(\beta\mathbf{z})$, where $\beta$ is a scale parameter that will be increased during training proceeds. Although HashNet is helpful in reducing the quantization error with scaled tanh function, it still needs sign function to generate exactly binary hash codes after training. 

The discrete optimization approach used in our work is related to STE~\cite{bengio2013estimating}, which can be seen as the base case of our proposed optimization strategy. STE uses sign function to generate exactly hash code in forward pass, and copies the gradient of $\mathbf{h}$ directly to $\mathbf{z}$ in back-propagation. It has been applied in many works~\cite{van2017neural,shen2018nash} to help the end-to-end learning of discrete codes. However, as we proved in section 4, standard STE is usually hard to train because of the noisy gradients contained in back-propagation due to the operation of gradient hard copy. Therefore, we adapt a guidance aware strategy to eliminate the influence of noisy gradients in STE and enable it can mimic the learning process of corresponding continuous embedding based optimization problem, in which faster training and better performance are achieved. 

\begin{figure*}[h]
  \centering
  \includegraphics[width=16cm,height=6.cm]{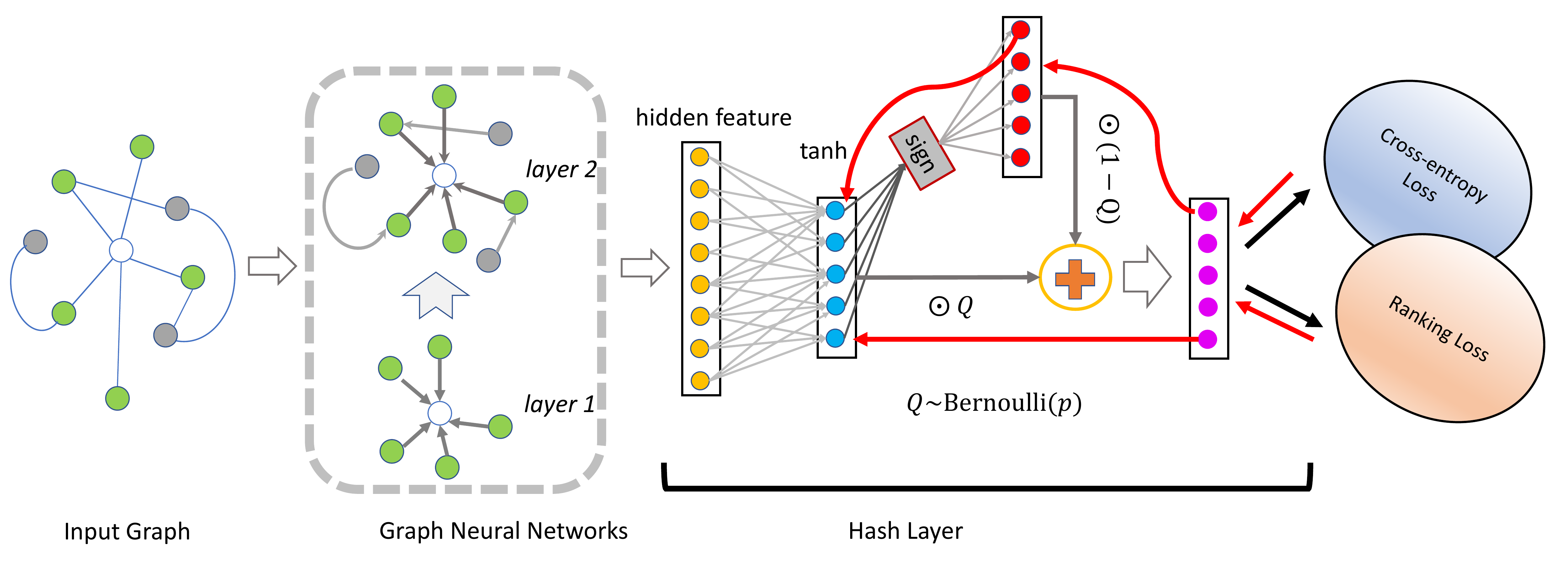}
  \caption{The architecture of HashGNN (better viewed in color). Our method consists of two main components: (1) a graph neural networks for learning deep representation of each node $v_i$, (2) a fully-connected hash layer with tanh activation for transforming the deep representation into $K$-dimensional continuous embedding $\mathbf{z}_i\in\mathcal{R}^K$, which is then fed into sign function to generate $K$-bit hash code $\mathbf{h}_i\in\{1,-1\}^K$. The whole architecture is trained end-to-end by jointly optimizing two losses, i.e., cross-entropy loss to recover the observed links and ranking loss to preserve the relative similarity ordering of hash codes. Bernoulli random variable $Q$ is introduced to provide average dropout of hash codes and continuous embeddings, which enables the discrete optimization problem is guided by continuous counterpart. Red arrows denotes gradient back-propagation.  }
  \label{figure2}
\end{figure*}

\section{The Proposed model}
In this section, we first introduce notations and define the research problem. Then, for our model, we describe the forward pass of graph encoder and hash layer for generating binary hash code. Finally, we introduce the ranking loss function for hashing.  

\subsection{Problem Formulation}
We are given a graph $\mathcal{G}=\{\mathcal{V},\mathcal{E},\mathbf{X}\}$ as input, where $\mathcal{V}$ denotes the set of vertices and $v\in\mathcal{V}$ represents a node in $\mathcal{V}$. $\mathcal{E}\subseteq \mathcal{V}\times{\mathcal{V}}$ denotes the edges. $\mathbf{A}\in\mathbb{R}^{N\times{N}}$ is the adjacency matrix, where $\mathbf{A}_{ij}=1$ denotes a link exists between $v_i$ and $v_j$, and otherwise $\mathbf{A}_{ij}=0$. $\mathbf{X}\in \mathbb{R}^{N\times D}$ is the associated attributes, $\mathbf{x}_i$ denotes the attributes of the $v_i$. In our settings of recommender systems, we focus on bipartite networks consisting of users and items. For simplicity, $v$ may denote either type of nodes in $\mathcal{G}$. The goal of our model is to learn a nonlinear hash function that maps a node $v$ to hamming space, i.e. $\mathcal{H}:\mathcal{V} \rightarrow \mathbf{h}\in\{-1, 1\}^K$. Specifically, we first build a graph encoder $\phi:\mathcal{V}\rightarrow \mathbf{u}\in\mathbb{R}^d$ to generate intermediate representation for the node. Then, the intermediate representation is fed into a hash layer $\psi:\mathbf{u}\rightarrow \mathbf{h}$ to learn binary hash codes. We denote $\Theta_{\phi}$ and $\Theta_{\psi}$ as trainable parameters for the graph encoder and hash layer, respectively. 

To measure the fitness of $\mathcal{H}$ in an unsupervised manner, we consider a differentiable computational graph that takes a pair of nodes $v_i$ and $v_j$ as inputs, and measures the probability that determines the existence of corresponding linkage.
Assuming that $\mathcal{L}$ denotes the loss function. The goal of obtaining $\{\mathbf{h}_i; i=1,...,N\}$ is to solve for $\Theta_{\phi}, \Theta_{\psi}$ via minimizing $\mathcal{L}(\mathbf{A}| \mathbf{X}, \Theta_{\phi}, \Theta_{\psi})$.

\subsection{Graph Encoder}
As introduced above, $\phi$ is a graph encoder that takes graph as input and outputs an intermediate vector $\mathbf{u}_i$ for any node $v_i$. In general, the choice of $\phi$ is very flexible and can be instantiated by any deep representation methods as discussed in~\cite{cui2018survey}. In this work, we consider a two-layer graph convolutional network GCN~\cite{hamilton2017inductive}. The major step of embedding generation in GCN is to aggregate information from a node's local neighbors. Let $\mathbf{u}^l_i$ denote the representation of node $v_i$ on layer of depth $l$, then the forward propagation rule is:
\begin{equation}
    \mathbf{u}^l_i = \sigma(\mathbf{W}^l\cdot \text{MEAN}\{\mathbf{u}^{l-1}_i\cup \{\mathbf{u}^{l-1}_j|\forall v_j\in\mathcal{N}(v_i)\}\}),
    \label{eq1}
\end{equation}
where $\mathbf{W}^l$ is the weight matrix on layer $l$, $\text{MEAN}$ refers to the average aggregator over feature vectors of the current node and its neighborhood nodes, and $\mathcal{N}(v_i)$ denotes the local neighborhood of $v_i$. The final output is $\mathbf{u}_i=\mathbf{u}^{l_{max}}_i$, where $l_{max}$ denotes the depth of final layer. 

\subsection{Hash Code Embedding}
In this section, we introduce the forward propagation of the hash layer to generate binary hash codes for nodes in $\mathcal{G}$. Given the intermediate representation $\mathbf{u}_i$ of node $v_i$, a fully-connected hash layer is adopted to transform $\mathbf{u}_i$ into $K$-dimensional embedding vector $\mathbf{z}_i\in\mathbb{R}^K$ as follows:
\begin{equation}
    \mathbf{z}_i = \sigma(\mathbf{W}^T\mathbf{u}_i + \mathbf{b}),
    \label{eq2}
\end{equation}
where $\mathbf{W}\in\mathbb{R}^{d\times{K}}$ is the parameter matrix, $\mathbf{b}\in\mathbb{R}^K$ is the bias matrix. To enforce the resultant embedding to be binary, the activation function $\sigma(\cdot)$ is usually set as $tanh$ function. In order to generate exactly binary hash code, we resort to convert the $K$-dimensional vector $\mathbf{z}_i$, which is continuous in nature, to binary code $\mathbf{h}_i$ with values of either $+1$ or $-1$ by taking the sign function, so that:
\begin{equation}
\mathbf{h}_i=sign(\mathbf{z}_i)=\left\{
\begin{aligned}
&+1, \ \text{if} \ \mathbf{z}_i \geq 0 \\
&-1, \ \text{otherwise} 
\end{aligned}
\right.
\label{eq3}
\end{equation}
Given the graph encoder and hash layer defined as above, a natural follow-up question is how to learn the parameters of our model. Different from classification learning tasks, where the label information is available to support supervised or semi-supervised learning, graph representation learning task often focuses on unsupervised learning, in which the goal is to generate similar vector embeddings for topologically similar nodes so that the graph structure could be preserved as much as possible. For each pair of node in $\mathcal{G}$, the likelihood function $P(\mathbf{A}_{ij}|\mathbf{h}_i, \mathbf{h}_j)$ can be interpreted as the conditional probability that the link connection between $v_i$ and $v_j$ equals $\mathbf{A}_{ij}$ given their hash codes $\mathbf{h}_i$ and $\mathbf{h}_j$, which could be naturally defined as pairwise logistic function as follows
\begin{equation}
P(\mathbf{A}_{ij}|\mathbf{h}_i, \mathbf{h}_j))=\left\{
\begin{aligned}
&\sigma(dist(\mathbf{h}_i, \mathbf{h}_j)), \ \ \mathbf{A}_{ij}=1 \\
&1 - \sigma(dist(\mathbf{h}_i, \mathbf{h}_j)), \ \  \mathbf{A}_{ij}=0,
\end{aligned}
\right.
\label{eq4}
\end{equation}
where $\sigma(\cdot)$ is the standard sigmoid function, $dist(\cdot,\cdot)$ represents the distance measure. Note that there exists a nice relationship between hamming distance $ham_{dist}(\cdot, \cdot)$ and inner product $\left\langle\cdot, \cdot\right\rangle$ for binary hash codes, where 
$ham_{dist}(\mathbf{h}_i, \mathbf{h}_j)=\frac{1}{2}(K-\left\langle\mathbf{h}_i, \mathbf{h}_j\right\rangle)$, thus we adopt inner product to define $dist(\cdot,\cdot)$. In this case, 
the smaller the hamming distance is, the larger the conditional probability $P(1|\mathbf{h}_i, \mathbf{h}_j)$ will be, and vice versa.
Therefore, Equation~\ref{eq4} is a reasonable extension of logistic regression classifier for the pairwise classification problem based on hash codes. 
Hence, we can achieve the optimal parameters when the following loss function is minimized:
\begin{equation}
\mathcal{L}_{cross}= -\sum_{\mathbf{A}_{ij}\in\mathbf{A}}\mathbf{A}_{ij}\log(\sigma(\left \langle\mathbf{h}_i, \mathbf{h}_j\right \rangle)) + (1-\mathbf{A}_{ij})\log(1 - \sigma(\left \langle\mathbf{h}_i, \mathbf{h}_j\right \rangle)), 
\label{eq5}
\end{equation}
where $\mathcal{L}_{cross}$ represents the cross-entropy loss that helps to reconstruct the observed links. $\Theta=\{\Theta_{\phi}, \Theta_{\psi}\}$ are the overall trainable parameters in our model. 
By minimizing Equation~\ref{eq5}, we obtain binary hash codes that share similar values between linked nodes, so that the network topological information can be stored in the hamming space.

\subsection{Ranking Preserving Hash Coding}
In general, Equation~\ref{eq5} helps in learning useful embedding vectors for nodes in graphs, and also ensure similar nodes to have similar representations. In recommender systems, however, relative rankings between candidate items are often more important than their absolute similarity scores~\cite{he2018adversarial, cao2018deep,park2015preference,zhao2016improving}, since recommendation engines return items mainly according to their rankings. For this reason, the learned hash codes in Equation~\ref{eq5} may lack the capability of mapping relevant objects to be within a small hamming ball, which is very important for high quality retrieval in the hamming space~\cite{cao2018deep}. In addition, since we target graph data, the ranking structure can be obtained easily in practice by sampling from the adjacency matrix. Inspired by this fact, we propose to preserve the relative similarity ordering of nodes in terms of the hamming space and introduce a ranking structure reinforced loss function $\mathcal{L}_{rank}$. Specifically, assuming that $(v_i,v_j,v_m)\in\mathcal{D}$ is a triplet denoting that $v_i$ is more similar to $v_j$ compared to $v_m$, and $\mathcal{D}$ is the set of all such sampled triplets. The ranking loss objective is defined as:
\begin{equation}
\mathcal{L}_{rank}=\sum_{(v_i,v_j,v_m)\in\mathcal{D}}\max(0, -\sigma(\left\langle\mathbf{h}_i,\mathbf{h}_j\right\rangle)+\sigma(\left\langle\mathbf{h}_i,\mathbf{h}_m\right\rangle) + \alpha), 
\label{eq6}
\end{equation}
where $\alpha$ is the margin parameter that helps control the difference between similar and dissimilar hash codes. By minimizing Equation~\ref{eq6}, the binary hash codes of neighbors for node $v_i$ will be concentrated in a small hamming radius near $v_i$ than nodes that are not connected with $v_i$. Hence, high quality hamming space retrieval can be achieved. We fix $\alpha$ to 0.2 in experiments.

\subsection{Recommendation Inference}
We now briefly introduce the schema of how to recommend relevant items given a query. After training HashGNN, we use the forward pass of the architecture to generate binary hash code $\mathbf{h}$ and continuous embedding $\mathbf{z}$ for each node in the input graph. Given a query $q$, assuming that $\mathbf{h}_q$ and $\mathbf{z}_q$ are the corresponding hash code and continuous embedding, respectively, we basically have two ways to do inference that trades off efficiency and effectiveness. The first is \textit{Hamming Space Retrieval}, in which a list of items are retrieved by sorting hamming distances of hash codes between the query and items in search pool. The second is \textit{Hierarchical Search}, in which it first retrieves a small set of candidates using hamming space retrieval, and then conduct the final recommendation by ranking the returned candidates with their continuous embeddings. The former strategy mainly focuses on efficiency, while the latter one aims to trade-off between efficiency and performance. 

\section{End-to-End Learning of Hash code}
By combining the considerations discussed above, the overall loss function to be minimized is formulated as
\begin{equation}
\begin{aligned}
\mathcal{L}&=\mathcal{L}_{cross} + \lambda \mathcal{L}_{rank}\\ &=-\sum_{\mathbf{A}_{ij}\in\mathbf{A}}\mathbf{A}_{ij}\log(\sigma(\left \langle\mathbf{h}_i, \mathbf{h}_j\right \rangle)) + (1-\mathbf{A}_{ij})\log(1 - \sigma(\left \langle\mathbf{h}_i, \mathbf{h}_j\right \rangle))\\
&+\lambda \sum_{(v_i,v_j,v_m)\in\mathcal{D}}\max(0, -\sigma(\left\langle\mathbf{h}_i,\mathbf{h}_j\right\rangle)+\sigma(\left\langle\mathbf{h}_i,\mathbf{h}_m\right\rangle) + \alpha),
\end{aligned}
\label{eq7}
\end{equation}
where $\lambda$ is the trade-off parameter to balance the importance between entropy and ranking loss. Note the combination of pair-wise ranking loss and point-wise reconstruction loss is usually helpful for recommendation~\cite{zhao2016improving}. 
Ideally, we hope the above loss function is differentiable w.r.t. parameters $\Theta=\{\Theta_{\phi}, \Theta_{\psi}\}$, so that it can be optimized through standard stochastic gradient descent or its variants~\cite{kingma2014adam}. Unfortunately, as the sign function is non-smoothing, its gradient is ill-defined as zero, making it apparently incompatible with back-propagation. 
In order to train our model in an end-to-end fashion, we approximate the gradient similar to straight through estimator STE~\cite{bengio2013estimating}. That is, we use sign function to generate hash code $\mathbf{h}=sign(\mathbf{z})$ in forward pass, where we copy the gradients from $\mathbf{h}$ directly to $\mathbf{z}$ in backward pass.  
With such gradient approximation, it is possible to optimize the hash model via back-propagation. 

\subsection{Learning with Guidance}
Although STE is simple to implement and has been widely applied for its effectiveness in approximating the gradient of discrete variables~\cite{van2017neural,shen2018nash}, it suffers from the gradient magnification problem as shown in Lemma 4.1, which makes the training unstable and may cause sub-optimal solutions. This phenomenon is similar to the well known \textit{gradient explosion} problem, which has been a key difficulty in training deep neural networks via back-propagation~\cite{lecun2015deep}.    

\medskip
\noindent\textbf{Lemma 4.1.} \textit{ Straight-through estimator (STE), which enables the optimization of discrete variables by copying their gradients directly to corresponding continuous variables, could result in sub-optimal solutions due to the magnification of gradients}. 

\smallskip
\noindent\textit{Proof}. Assuming that $\mathcal{H}$ is an arbitrary graph neural network with tanh activation function in the final layer. It takes a node $v$ as input and outputs an embedding denoted as $\mathbf{z}$. If we consider cross entropy loss, then for a pair of nodes $(v_i, v_j)$, we have $\mathcal{L}(y_{ij}|v_i, v_j)=y_{ij}\log(\sigma(\hat{y}_{ij}))+(1-y_{ij})(1-\log(\sigma(\hat{y}_{ij})))$, where $\hat{y}_{ij}=\mathbf{z}_i\mathbf{z}_j^T$ and $y_{ij}$ is the ground truth. Let $g_{\hat{y}_{ij}}$ denotes the gradient $\frac{\partial \mathcal{L}}{\partial \hat{y}_{ij}}$, we have $g_{\mathbf{z}_{ij}}=g_{\hat{y}_{ij}}\mathbf{z}_j$ for standard unsupervised graph representation learning. Now if we have the sign function on the top of $\mathcal{H}$ that helps generate binary hash code denoted as $\mathbf{h}$, and want to learn the hash code end-to-end, that is $\hat{y}_{ij}=\mathbf{h}_i\mathbf{h}_j^T$. In this case, it is easily obtain the new gradient denoted as  $g_{\mathbf{z}_i}^{ste}=g_{\mathbf{h}_i}^{ste}=g_{\hat{y}_{ij}\mathbf{h}_j}$. We can see that $g_{\mathbf{z}_i}^{ste}$ differs from $g_{\mathbf{z}_{ij}}$ in the lash term, $\mathbf{h}_j$ for $g_{\mathbf{z}_i}^{ste}$ while $\mathbf{z}_j$ for $g_{\mathbf{z}_{ij}}$. Recall that $\mathbf{h}_i=sign(\mathbf{z}_i)$ and $\mathbf{z}_i$ is the output of $tanh$ function, then $\forall m$, $\mathbf{z}_{jm}\leq\mathbf{h}_{jm}$, Hence, the optimization problem of STE is similar to standard graph representation learning excepting the magnification of gradients with binarization. The gradient magnification of STE could impede solving for the optimal parameters that standard representation learning could achieve, because STE consistently takes a larger step in searching optimal gradients.
\medskip

Based on the above discussion, we know that STE actually optimizes a similar problem with standard representation learning methods, except the magnification of gradients in each step. 
Since graph representation learning approaches are usually easy to train and can generate high quality embedding vectors~\cite{cai2018comprehensive}, a natural question is can we assist the optimization process of STE with the help of corresponding continuous embedding? Motivated by their connections, we propose to associate the regular continuous embedding vector $\mathbf{z}_i$ during training, and make the exactly hash codes mimic the corresponding continuous embedding vectors, so that both of them are simultaneously optimized for the task. More specifically, during training, instead of using the generated binary hash codes, we use a dropout average of them as follows,
\begin{equation}
\hat{\mathbf{h}}_i=Q\odot\mathbf{z}_i + (1-Q)\odot\mathbf{h}_i, 
\label{eq8}
\end{equation}
where $Q$ represents a Bernoulli random variable with parameter $p$ for selecting between regular continuous embedding vectors and the exactly hash codes. 
When $p$ is set to a relatively high probability, even if $\mathbf{h}_i$ is difficulty to learn, $\mathbf{z}_i$ can still be learned to assist the improvement of the task-specific parameters, which in turn helps hash code generation. In addition, the guidance formulation above can be seen as a variant of Dropout~\cite{srivastava2014dropout}. Instead of randomly setting a portion of the activations in $\mathbf{h}$ to zeros when computing the parameters gradients, we update them with the continuous counterparts, which can help to generalize better.  

In experiments, instead of setting $p$ with a fixed value by hyperparameter tuning, we use a large $p$ at early iterations, and then gradually decrease $p$ as training proceeds, as inspired by recent studies in continuation methods~\cite{allgower2012numerical}. By gradually reducing $p$ during training, it results in a sequence of optimization problems converging to the original optimization problem of STE. As shown in experiments, it improves both training process and testing performance.

\section{Experiments}
In this section, we conduct experiments over three recommendation datasets to validate the proposed approach. Specifically, we try to answer the following  questions: 
\begin{itemize}
    \item \textbf{Q1}: How effective is the proposed method compared to baselines in the hamming space retrieval?
    \item \textbf{Q2}: How does HashGNN perform compared with state-of-the-art deep hashing methods in hierarchical retrieval search tasks?
    \item \textbf{Q3}: What the benefits of STE with continuous embedding guidance compared with original STE?
    \item \textbf{Q4}: How do different hyper-parameter settings (e.g., trade-off parameter $\lambda$, number of triplets for each node, embedding dimension $K$) affect HashGNN?
\end{itemize}

\begin{figure*}[h]
  \centering
  \includegraphics[width=18cm,height=5cm]{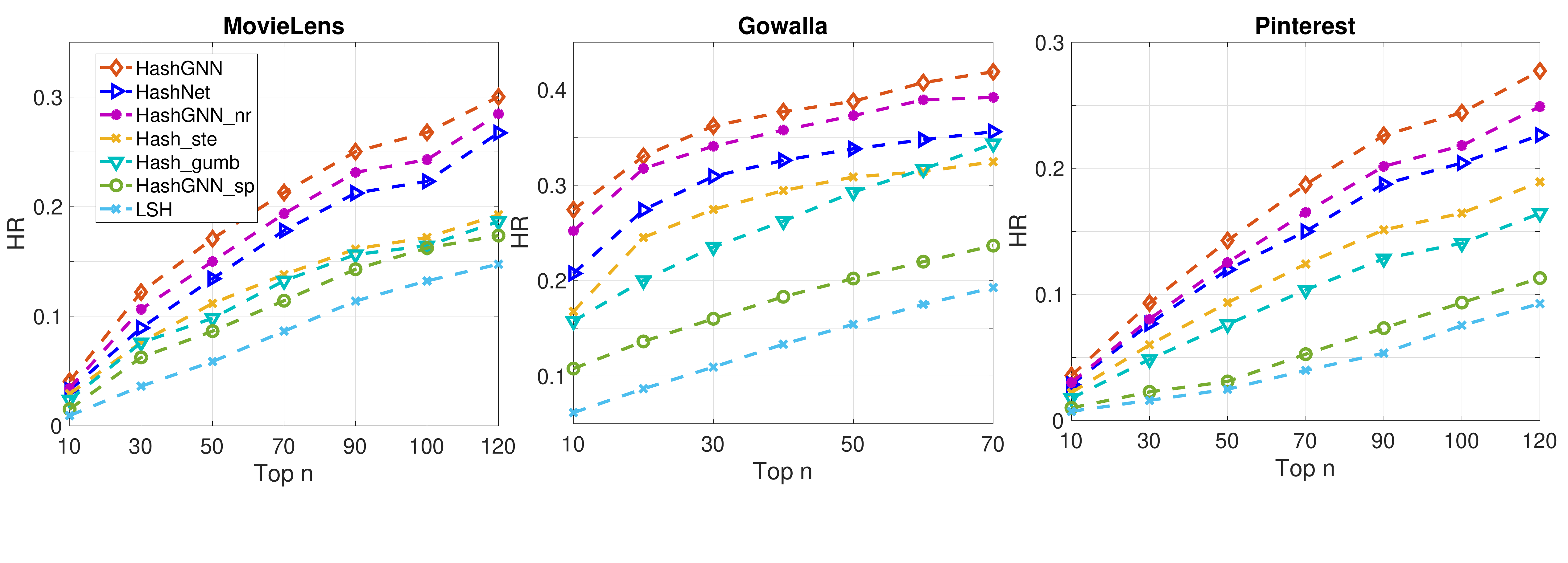}
  \caption{Recommendation performance comparison for Hamming Space Retrieval.}
  \label{figure3}
\end{figure*}

\begin{table}
  \caption{Statistics of the datasets.}
  \begin{tabular}{c|c|c|c}
    \toprule
    \hline
    Dataset &\#Users &\#Items &\#Interactions\\
    \hline
    \midrule
    MovieLens & 6,040 &  3,952 &1,000,209\\
    \hline
    Gowalla &29,585 &40,981 &1,027,370\\
    \hline
    Pinterest &55,187  &9,916 &1,500,809\\
    \hline
    Alibaba &601,946  &4,167,708 &28,489,957\\
   \hline
  \bottomrule
\end{tabular}
\label{table1}
\end{table}
\subsection{Datasets}
To evaluate the effectiveness of HashGNN, we conduct experiments on four benchmark datasets, some of publicly accessible and differ in terms of domain and size. We summarize the statistics of three datasets in Table 1.

\noindent \textbf{Gowalla}: This is the check-in dataset~\cite{liang2016modeling} obtained from Gowalla, in which users share their locations by check-in. To ensure the
quality of the dataset, we retain users and items with at least ten interactions similar to~\cite{wang2019neural}.

\noindent \textbf{MovieLens}: This is widely used movie rating dataset~\footnote{https://grouplens.org/datasets/movielens/1m/}. We treat it as a heterogeneous networks where links exist between users and movies. Similar to~\cite{he2016fast}, we transform the rating scores into binary values, so that each entry is either 1 or 0 indicating whether the user rated the movie. The 1M version is used in our experiments. 

\noindent \textbf{Pinterest}: This is an implicit feedback data constructed for image recommendation~\cite{geng2015learning}. We treat users and images as nodes. The link represents a pin on an image initiated by a user. Each user has at least 20 links. 

\noindent \textbf{Alibaba}: This dataset is a subset of user behavior data in A+ company. It describes user's historical behaviors including click, chart and purchase toward items during September 9 to September 15, 2019. The data is organized similar to MovieLens dataset, i.e., a user-item behavior consists of user ID, item ID, item's, behavior type and timestamp. We filter the users that has less than 10 interactions for experiments and use the former six consecutive days for training while the last day for testing. 

For the previous three datasets, we randomly select 70\% of historical interactions of each user to constitute the training set, and treat the remaining as test set. From the training set, we randomly
select 10\% of interactions as validation set to tune hyper-parameters.
For each observed user-item interaction, we treat it as a positive instance, and then construct ranking triplets by sampling from negative items that the user did not consume before.

\subsection{Experimental Settings}
\subsubsection{\textbf{Evaluation Metrics.}} For each user in the test set, we treat all the items that the user has not interacted with as the negative items. To evaluate the performance of top-$n$ recommendations, we adopt two widely used evaluation metrics~\cite{he2017neural}: \textit{hit rate} (HR) and \textit{normalized discounted cumulative gain} (ndcg) over varying numbers of top returned items. The average results over all users in the test set are reported.

\begin{table*}[h]
\small
  \caption{Recommendation performance comparison for Hierarchical Search.}
  \begin{tabular}{l|cccc|cccc|cccc}
    \toprule
    \multirow{2}*{} &\multicolumn{4}{c}{\textbf{MovieLens}} &\multicolumn{4}{c}{\textbf{Gowalla}} &\multicolumn{4}{c}{\textbf{Pinterest}} \\
   \cmidrule(lr){2-5}\cmidrule(lr){6-9} \cmidrule(lr){10-13}
    &HR@50 &HR@100 &ndcg@50 &ndc@100 &HR@50 &HR@100 &ndcg@50 &ndc@100 &HR@50 &HR@100 &ndcg@50 &ndc@100\\
    \midrule
    \textbf{LSH} &0.063 &0.127 &0.143 &0.192 &0.165 &0.235 &0.282 &0.378 &0.044 &0.093 &0.066 &0.117\\
    \textbf{HashGNN\_sp} &0.108 &0.177 &0.207 &0.272 &0.219 &0.261 &0.375 &0.443 &0.052 &0.113 &0.083 &0.127\\
    \textbf{Hash\_gumb} &0.136 &0.220 &0.234 &0.324 &0.264 &0.335 &0.452 &0.507 &0.098 &0.172 &0.104 &0.149\\
    \textbf{Hash\_ste} &0.145 &0.225 &0.249 &0.335 &0.331 &0.363 &0.485 &0.516 &0.126 &0.208 &0.127 &0.175\\
   
    \textbf{HashNet} &0.185 &0.216 &0.266 &0.372 &0.346 &0.388 &0.502 &0.530 &0.169 &0.282 &0.147 &0.188\\
    \textbf{HashGNN\_nr} &0.223 &0.340 &0.294 &0.405 &0.378 &0.423 &0.533 &0.557 &0.189 &0.315 &0.162 &0.204\\
    \hline
    \textbf{MF} &0.187 &0.256 &0.276 &0.383 &0.345 &0.369 &0.504 &0.523 &0.041 &0.072 &0.104 &0.152\\
    \textbf{PTE} &0.159 &0.249 &0.257 &0.353 &0.344 &0.356 &0.498 &0.517 &0.031 &0.052 &0.088 &0.125\\
    \textbf{BiNE} &0.209 &0.289 &0.283 &0.392 &0.363 &0.389 &0.515 &0.546 &0.109 &0.248 &0.117 &0.186\\
    \textbf{GraphSage} &0.228 &0.354 &0.304 &0.411 &0.389 &0.434 &0.542 &0.563 &0.193 &0.317 &0.166 &0.213\\
    \hline
    \textbf{HashGNN} &\textbf{0.248} &\textbf{0.373} &\textbf{0.325} &\textbf{0.431} &\textbf{0.405} &\textbf{0.443} &\textbf{0.551} &\textbf{0.572} &\textbf{0.212} &\textbf{0.334} &\textbf{0.176} &\textbf{0.234}\\
  \bottomrule
\end{tabular}
\label{table2}
\end{table*}

\subsubsection{\textbf{Baselines.}} To the best of our knowledge, these is no existing work that directly study the problem of unsupervised deep hashing with graph neural networks, so we adapt several baselines that have shown strong performance on image retrieval task for comparison. In addition, we also include several state-of-the-art recommendation methods that based on network embedding.
\begin{itemize}
\item \textbf{PTE}~\cite{tang2015pte}, \textbf{BiNE}~\cite{gao2018bine}, \textbf{MF}~\cite{luo2014efficient} and \textbf{Graphsage}~\cite{hamilton2017inductive}: They are commonly used embedding methods for recommendation. Here we build our proposed hashing model based on GraphSage for its scalability and deep expressive ability. 

\item \textbf{LSH}~\cite{gionis1999similarity}: This is a state-of-the-art unsupervised hashing method. For fair comparison, the input feature is the output of Graphsage. 

\item \textbf{HashNet}~\cite{cao2017hashnet}: This is the state-of-the-art deep hashing method based on continuation methods~\cite{allgower2012numerical}. We adapt it for graph data by replacing the original AlexNet~\cite{krizhevsky2012imagenet} architecture with graph neural networks. 

\item \textbf{Hash\_gumb}: This is the implementation of learning to hashing with Gumbel-softmax~\cite{jang2016categorical}. The basic idea is to treat each bit element as a one-hot vector of size two. In the top of hash layer, we use Gumbel-softmax trick instead of $sign$ function to generate hash codes. Note that by gradually decreasing the temperature parameter towards zero, it can obtain exactly binary hash codes in theory.  

\item \textbf{Hash\_ste}: This is the implementation of state-of-the-art end-to-end hash learning method based on straight through estimator~\cite{bengio2013estimating}. It is also a special case of our model. 

\item \textbf{HashGNN\_sp}: This is a variant of the proposed method by separating the graph embedding generation process with hash function learning. It first trains Graphsage to obtain node embeddings. The learned embeddings are then used as input feature to train the hash layer. 

\item \textbf{HashGNN\_nr}: This is a variant of the proposed method by excluding ranking loss. We introduce this variant to help investigate the benefit of using triplet loss.
\end{itemize}

\subsubsection{\textbf{Parameter Settings}.} For fair comparison, all methods are implemented in Tensorflow. We adopt a two layer graph convolutional networks~\cite{hamilton2017inductive} for graph encoding, where the embedding size for first and second layer are fixed as 128 and 68 for all datasets, respectively. We optimize all models with the Adam optimizer with a mini-batch size of 256. All weights are initialized from a zero-centered Normal distribution with standard deviation 0.02, and the learning rate is fixed as 0.001. For HashNet, the scale parameter $\beta$ for $tanh(x)=(\beta x)$ is initially set as 1, and then it increases exponential after 200 iterations as suggested in ~\cite{cao2017hashnet}. The temperature for Hash\_gumb is implemented with initial value 1 by tf.contrib.distributions.RelaxedOneHotCategorical according to~\cite{jang2016categorical}. For network embedding approaches, we use the default parameter used in original papers. In our method, the trade off parameter $\lambda$ is tuned from 0.1 to 1 with step size 0.1; parameter $p$ for Bernoulli distribution is started with 1 and then we decrease it 5\% after 250 iterations on all datasets. Without specification, the default embedding size is 32. 

\subsection{Performance Comparison on Hamming Space Retrieval (Q1)}
In this section, we evaluate the effectiveness of our proposed method in generating hash codes. Specifically, the retrieved products for each user are obtained by sorting hamming distances of hash codes between each user and all candidate items. Figure~\ref{figure3} reports the results in terms of HR for different top-$n$ values. The results on ndcg are similar, we omit them for space limitation. 

From the figure, we have four observations. First, LSH separates the encoding of feature representation from hashing and it achieves poor performance on three datasets. This indicates the importance of jointly learning feature representation and hash projections for high quality hash code generation. Second, HashNet and Hash\_gumb are two widely used continuous relaxation based approaches for hash learning. But HashNet outperforms Hash\_gumb in several cases. A possible reason is HashNet adopts continuation methods to train the model, which approximates the original hard optimization problem with sign function with a sequence of soft optimization problems, making it easier to learn. Third, Compared to HashNet, although Hash\_ste is a truly end-to-end hashing method, it is still outperformed by HashNet in most cases. It makes sense since Hash\_ste will magnify the gradients, which makes it deviated from standard continuous embedding learning problem. This demonstrates our motivation to guide ste with continuous embedding. Fourth, In general, HashGNN consistently yields the best performance on all the datasets. Increasing the number of $n$ has positive influence on the performance of all methods. By using continuous embedding guidance, HashGNN is capable of mimicking the learning process of continuous embedding optimization, while ste deviates from continuous embedding learning problem by magnifying gradients. This verifies the effectiveness of continuous embedding guidance for hashing. Moreover, compared with HashGNN\_nr, HashGNN considers ranking loss to preserve the relative similarity ordering of nodes in hamming space, while HashGNN\_nr only targets to reconstruct the observed links. This demonstrates the importance of capturing ranking structure in hamming space. And the improvements over HashGNN\_sp indicate that jointly optimizing feature representations and hash functions can achieve better performance. 

\subsection{Performance Comparison on Hierarchical Search (Q2)}
In this section, we attempt to understand the representative ability of node embeddings that learned by hashing methods as well as continuous network embedding approaches. Towards this end, 
we conduct experiments on three datasets and evaluate the performance based on hierarchical search. 
Different from hamming space search, hierarchical search needs both hash codes and continuous embeddings for inference. Therefore, it measures the ability of hashing models in generating high quality binary codes as well as good continuous embeddings. The results are shown in Table~\ref{table2}. 

From Table~\ref{table2}, we have four findings. First, Compared with hashing approaches, HashGNN\_nr performs significantly better than others almost in all cases. These results demonstrate the effectiveness of the proposed guidance learning approach for optimizing discrete variables. Second, Compared with continuous network embedding methods, HashGNN\_nr still achieves relative better performance than other methods except GraphSage. It is reasonable as HashGNN\_nr is build upon powerful graph neural networks (GraphSage), which outperforms other baselines significantly. These results illustrate our motivation to build hashing alternatives for arbitrary graph neural networks. Third, By jointly analyzing Table~\ref{table2} and Figure~\ref{figure3}, we observe that the performances of all hashing baselines in hierarchical search scenario are higher than their performance on hamming space retrieval. This indicates the limitation of hash codes for accurate retrieval compared with continuous embeddings. Hence, hashing models should also consider how to generate representative continuous embeddings in order to achieve satisfactory performance. Fourth, HashGNN consistently outperforms all other baselines on three datasets. Specifically, HashGNN achieves better performance than hashing methods in most cases, which demonstrates the effectiveness of our method in generating high quality hash codes as well as learning representative node embeddings. Compared with HashGNN\_nr, HashGNN is capable of preserving relative ranking structure in hamming space, while HashGNN\_nr only focuses on reconstructing the observed links. This validates that exploring ranking structure is beneficial for recommendation tasks. 

\begin{table}[h]
  \caption{Top $n$ recommendation performance on real-world Alibaba company dataset.}
  \begin{tabular}{l|cccc}
    \toprule
    \multirow{2}*{} 
    &\multicolumn{4}{c}{\textbf{Alibaba}} \\
   \cmidrule(lr){2-5}
     &HR@50 &HR@100 &ndcg@50 &ndc@100\\
    \midrule
    \textbf{LSH} &0.72\% &3.65\% &1.02\% &2.98\%\\
    \textbf{HashGNN\_sp} &1.17\% &5.51\% &1.97\% &6.59\%\\
    \textbf{Hash\_gumb} &1.81\% &5.90\% &2.26\% &7.14\% \\
    \textbf{Hash\_ste} &2.87\% &6.78\% &3.83\% &8.59\%\\
    \textbf{HashNet} &3.63\% &7.37\% &5.38\% &9.30\%\\
    \textbf{HashGNN\_nr} &4.48\% &9.30\% &6.65\% &11.18\%\\
    \hline
    \textbf{MF} &3.31\%  &7.22\% &5.18\% &9.19\%\\
    \textbf{PTE} &3.11\% &7.08\% &4.84\% &9.03\%\\
    \textbf{BiNE} &3.94\% &8.29\% &5.89\% &10.22\% \\
    \textbf{GraphSage} &4.73\% &9.86\% &6.91\% &11.44\%\\
    \hline
    \textbf{HashGNN} &\textbf{5.17\%}  &\textbf{10.66\%}  &\textbf{7.75\%}  &\textbf{12.84\%} \\
  \bottomrule
\end{tabular}
\label{table3}
\end{table}
\begin{table}
  \caption{Efficiency comparison on all datasets (in seconds). The time cost consists of computing and sorting the distances. CeS denotes continuous embedding based search, HieS represents hierarchical search, and HamR denotes hamming space search.}
  \begin{tabular}{c|c|c|c|c}
    \toprule
     &MovieLens &Pinterest &Gowalla &Alibaba\\
    \hline
    \midrule
    CeS & 3.8 &  73.5 &189.6 &15181.7\\
    HieS &2.8 ($\times$ \textbf{1.3}) &41.2 ($\times$ \textbf{1.7}) &82.1 ($\times$ \textbf{2.3}) &4295.8 ($\times$ \textbf{3.5})\\
    HamR &1.5 ($\times$ \textbf{2.5}) &27.8 ($\times$ \textbf{2.6}) &66.1 ($\times$ \textbf{2.8}) &4163.9 ($\times$ \textbf{3.6})\\
  \bottomrule
\end{tabular}
\label{table4}
\end{table}

In addition to the three publicly accessible small datasets, we also test the performance of our model on real-world recommendation scenarios over Alibaba dataset. We adopt the same experimental configurations as above, and report the results in Table~\ref{table3}. 

From the table, we have three observations. First, The performance of all methods in terms of HR and NDCG values significantly lower than their values on MovieLens, Gowalla and Pinterest, while HashGNN still consistently outperforms both hashing and continuous based baselines. Second, compared with hashing methods, HashGNN\_nr achieves better performance than them in almost all cases including Hash\_ste. This result demonstrate the effectiveness of the proposed guidance aware STE optimization strategy. In addition, HashGNN\_nr also performs better than most network embedding approaches, i.e., MF, PTE and BiNE. The main reason is that our model is build upon powerful graph neural networks given the big performance gap between GraphSage and other baselines. These results validate our motivation to develop hashing methods for arbitrary graph neural networks. Third, although GraphSage performs slightly better than HashGNN\_nr, it is outperformed by HashGNN in most cases. This comparison not only demonstrates our model could achieve comparable performance with its continuous counterpart, but also indicates that hashing methods can performs even better by considering the ranking structure in hamming space. 

Besides, we also analyze the efficiency of hierarchical search compared with hamming space retrieval and continuous embedding based search. For fair comparison, we adopt exhaustively linear scan to retrieval items for hash codes and continuous embeddings. It is worth noting that multi-level indexing techniques~\cite{norouzi2012fast} could be used to further improve the efficiency of hamming space retrieval. The experiments are conducted on MacOS 10.13.6 with Inter Core i5, 16GB RAM and python 3.6. Table~\ref{table3} summarized the results. 

From the table, we have two conclusions. First, hamming space retrieval is generally faster than other two approaches, while hierarchical search is more efficient than continuous embedding based retrieval. By jointly analyzing Table~\ref{table2}$, \ref{table3}\&$\ref{table4}, and Figure~\ref{figure3}, we can see that HashGNN not only can achieve better performance than GraphSage on hierarchical search, but also saves a lot of time (on average of \textbf{2.2} times faster). It thus verifies our motivation to combine graph representation learning with hashing. Second, as data size increases, the time cost of hierarchical search becomes closer to that of hamming space retrieval compared with linear search on continuous embedding space. The is mainly because hamming space retrieval becomes the most time-consuming step of hierarchical search in large datasets, since the second step only needs to compute similarity with very small set of items. This observation validates the capacity of end-to-end hashing in handling real-world large scale data.

\begin{figure}[h]
  \centering
  \includegraphics[width=9cm,height=4cm]{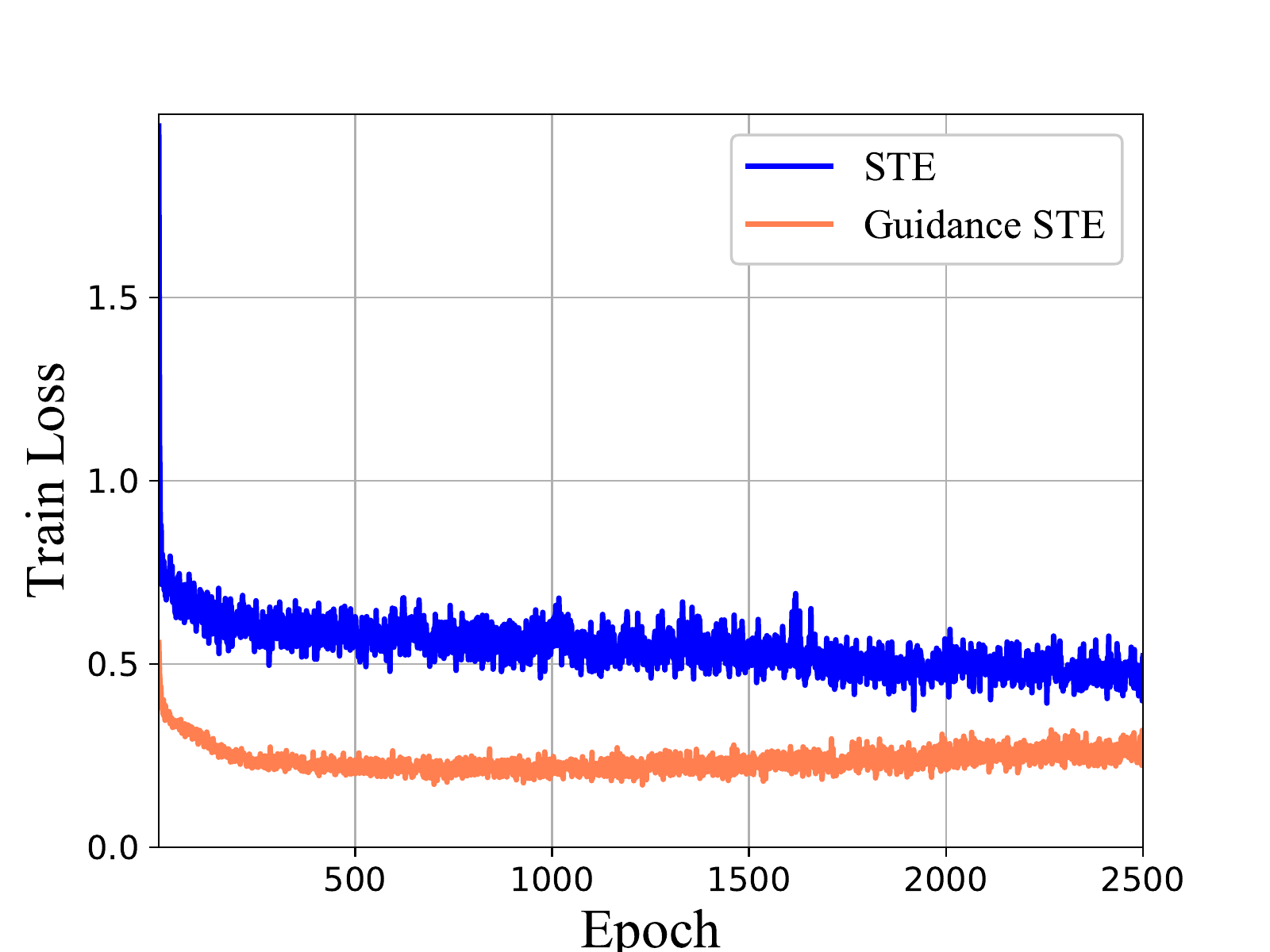}
  \caption{Guidance STE VS. STE over training loss.}
  \label{figure4}
\end{figure}

\begin{figure*}[h]
  \centering
  \includegraphics[width=14cm,height=5cm]{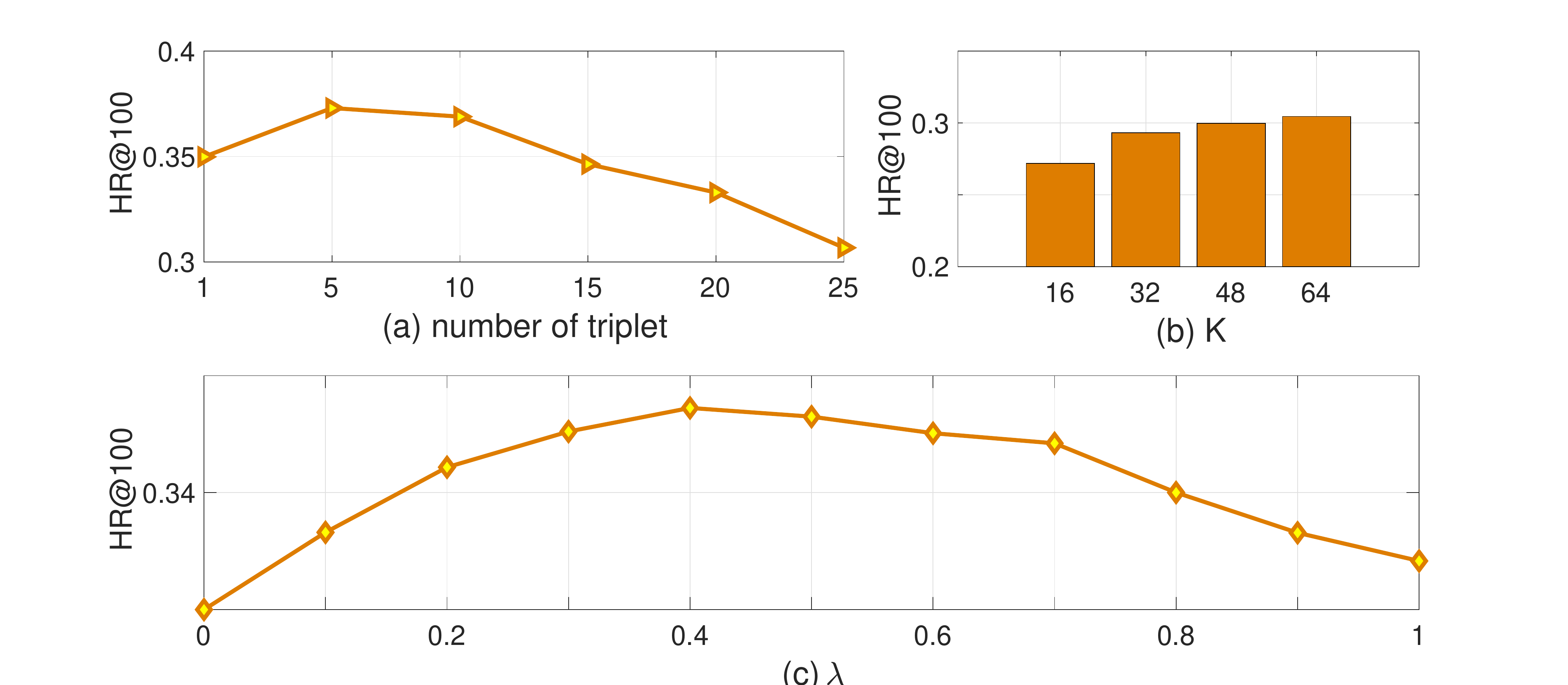}
  \caption{Parameter analysis of HashGNN w.r.t. number of triplet, $\lambda$ and $K$ on hierarchical search.}
  \label{figure7}
\end{figure*}

\subsection{Study of HashGNN (Q3) }
\begin{figure}[h]
  \centering
  \includegraphics[width=8cm,height=4.cm]{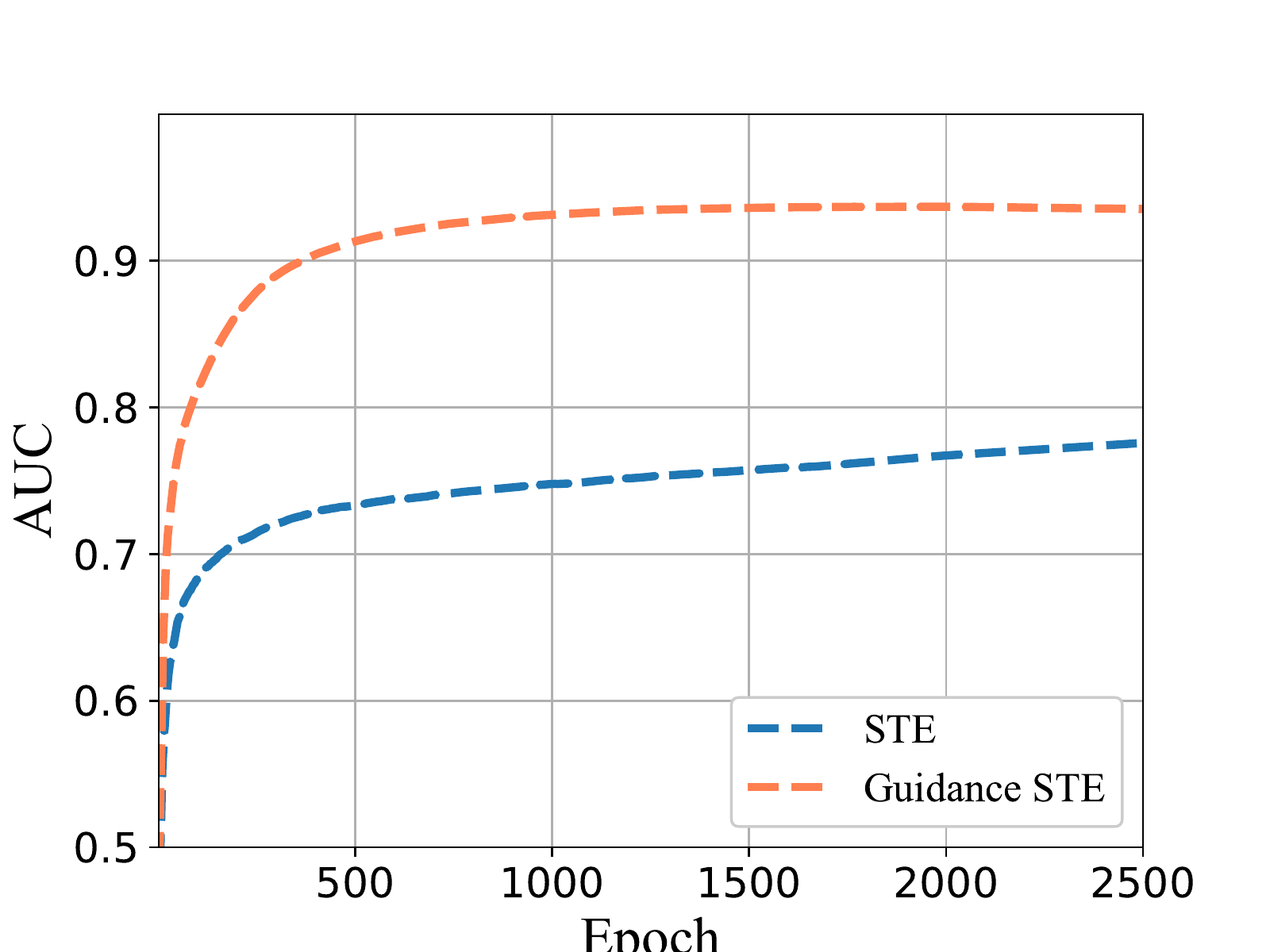}
  \caption{Guidance STE VS. STE over AUC.}
  \label{figure5}
\end{figure}

In this section, we empirically investigate the benefits of the proposed guidance aware STE optimization strategy. For fair comparison, we compare the performance of Hash\_ste and HashGNN\_nr, where Hash\_ste utilizes STE for discrete optimization while HashGNN\_nr exploits continuous embedding guidance STE. Similar to default setting, we set $p=1$ and gradually decrease it 5\% after 250 iterations. Figure~\ref{figure4} and~\ref{figure5} summarized the training loss and AUC scores during training on Alibaba datasets over hierarchical search, respectively. 

From the figures, we have three observations. First, the error loss of Guidance STE decreases faster than STE, and is also significantly lower than that of STE in general as shown in Figure~\ref{figure4}. This is reasonable because guidance STE solves the original hard discrete optimization problem by dividing it into a sequence of relative easier tasks, making it easier to optimize. This result indicates the efficiency of guidance STE over STE. Second, after rapid decrease of training loss, both guidance STE and STE begin to jitter, but guidance STE is more stable than STE. Besides, the error loss value of guidance STE also tends to increase a litter bit after 2000 epochs. This is mainly because the $p$ value of guidance STE decreases to 0.5, making the discrete optimization problem becomes more difficulty to optimize.  Third, guidance STE substantially outperforms STE in terms of AUC during training shown in Figure~\ref{figure5}. The above comparisons demonstrate the effectiveness and efficiency of the proposed guidance STE compared with STE.

\begin{figure}[h]
  \centering
  \includegraphics[width=9cm,height=4.cm]{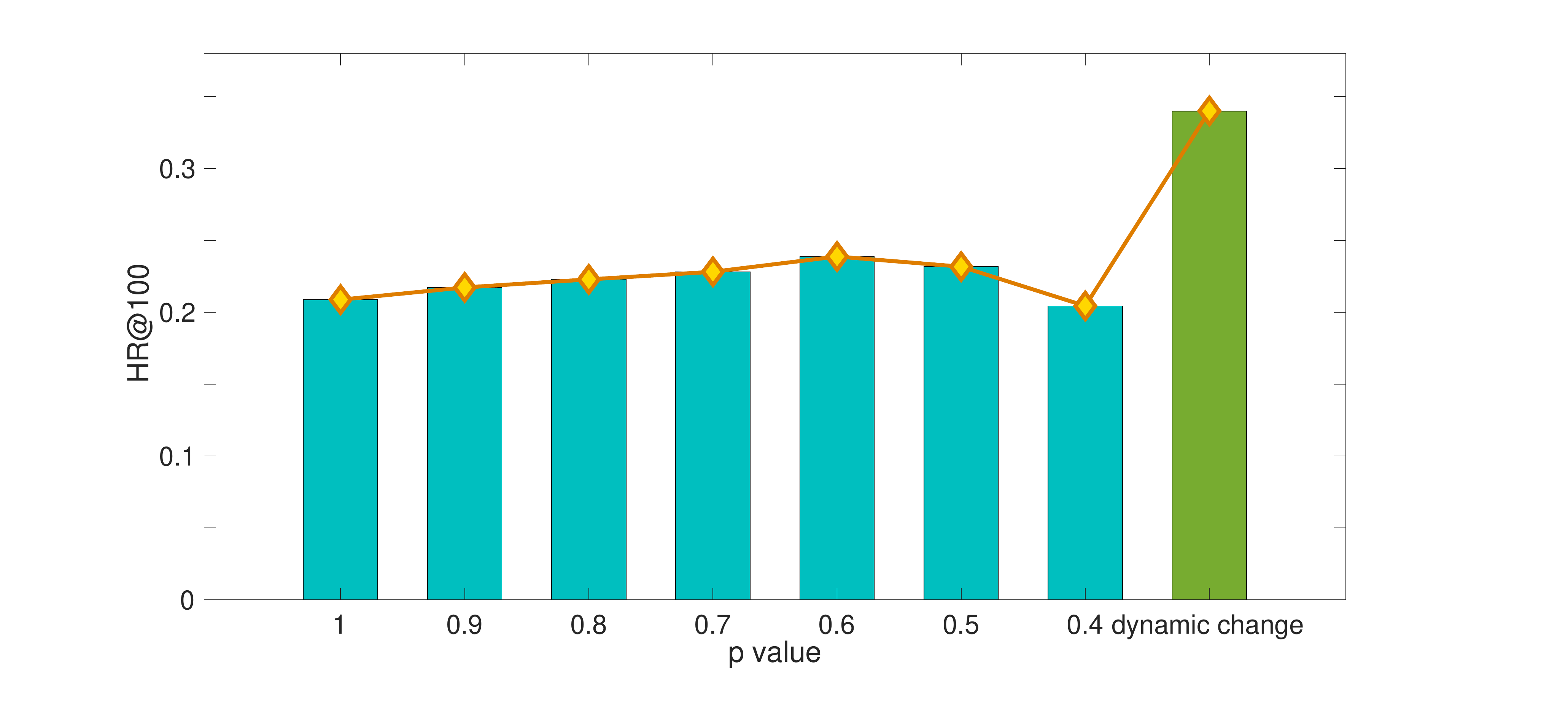}
  \caption{Impact of $p$. The last column of the bar chart denotes the performance by changing the $p$ value dynamically.}
  \label{figure6}
\end{figure}

Besides making comparisons to STE, we also analyze the impact of $p$ towards guidance STE. The experiment results are shown in Figure~\ref{figure6}. Results on other datasets are similar. Note that when $p=1$, guidance STE is equivalent to GraphSage. From the figure, we have two findings. First, the performance increases when $p$ varies from 1 to 0.6, then it tends to decrease from 0.6 to 0.4, and the HR score in 0.4 is lower than that when $p=1$. The possible reason is that when $p=1$, the objective is only targeted to generate high quality continuous embedding; while $p$ decreases, the objective function is optimized towards both hash code and continuous embedding, which improves the hierarchical search performance; when $p$ is too small, i.e., $p <0.5$, the continuous embedding signals are not enough to guide the training. 
Second, dynamically changing $p$ performs substantially better than $p$ is fixed. This this because when $p$ is dynamically updated, the original hard discrete optimization problem will be divided into a sequence of sub-tasks, where each task is generally easier to solve compared with the original one. In addition, each previous task in the sequence could also help solve the followed optimization tasks. Therefore, we change $p$ dynamically in our experiments. 

\subsection{Parameter Analysis (Q4)}
In this experiment, we study the effect of different hyperparameters to HashGNN. As $\lambda$ plays a pivotal role to trade-off between reconstruction loss and ranking preserving loss, we start by exploring the influence of $\lambda$. Then, we study how the number of triplets of each user affects the performance. Note we report results on MovieLens dataset, but similar results are observed on others. 

\subsubsection{\textbf{Effect of trade-off parameter}} To investigate the sensitivity of HashGNN w.r.t. $\lambda$, we search $\lambda$ from 0 to 1 with step size 0.1. Figure~\ref{figure4}c summarizes the experimental results on MovieLens, where we have two observations. First, HashGNN achieves relative stable performance when $\lambda$ is around 0.5, so we let $\lambda=0.5$ in our experiments. Second, when $\lambda=0$, HashGNN performs the worst. The reason is that when $\lambda=0$, the ranking loss is totally ignored. It thus further demonstrates our motivation to incorporate ranking loss for effective hashing.

\subsubsection{\textbf{Effect of the Number of Triplets}} To study the influence of different number of triplets on our model. We conduct additional experiments by varying the number of triplets for each node from 1 to 25 with step size 5. Figure~\ref{figure4}a reports the result on MovieLens. Results on other datasets are similar, so we omit them. 
From the figure, we can observe that HashGNN achieves relative stable and high performance when the number of triplets is between 4 to 10. 
Therefore, we randomly sample five triplets for each node in experiments. 

\subsubsection{\textbf{Effect of Embedding Dimension}} We now explore how the embedding size of hash code affects HashGNN. Figure~\ref{figure4}b lists the performance of HashGNN with different $K$ on MovieLens. We can observe that the dimension of hash code has positive influence on HashGNN. Specifically, HashGNN improves faster when $K$ is small, then it slows down with large dimension. Similar trends are observed on other datasets. 

\section{Conclusion and future Work}
In this work, we study the problem of unsupervised deep hashing with graph neural networks for recommendation.
We propose a new framework HashGNN, which simultaneously learning deep hash functions and graph representations in an end-to-end fashion. The proposed method is flexible as it can be used to extend existing various graph neural networks. The whole architecture is trained by jointly optimizing two losses, i.e., reconstruction loss to reconstruct the observed links and ranking preserving loss to preserve the relative similarity ranking of hash codes. To enabling the gradient propagation in the backward pass, we prove that the established straight through estimator that approximates the gradient of $sign$ function as identity mapping may result in noisy gradients, so we derive a novel discrete optimization strategy based on straight through estimator with continuous guidance. We show the proposed guidance is helpful in accelerating the training process as well as improving the performance. Extensive experiments on four real-world datasets demonstrate the effectiveness and efficiency of HashGNN in hamming space retrieval and hierarchical search. 

For future work, we plan to further improve HashGNN by incorporating semantic information of graph. Moreover, we are interested in exploring the adversarial learning on binary continuous embedding for enhancing the robustness of HashGNN. Finally, we are also interested in exploring the explainability of hash codes.

\section{ACKNOWLEDGMENTS}
The authors thank the anonymous reviewers for their helpful comments. The work is in part supported by NSF IIS-1718840, IIS-1750074, and IIS-1900990. The views and
conclusions contained in this paper are those of the authors and
should not be interpreted as representing any funding agencies.

\newpage
\bibliographystyle{ACM-Reference-Format}
\bibliography{ppr}

\appendix


\end{document}